\newif\ifdouble
\doubletrue

\ifdouble 
\documentclass[editor]{MetroTeX}
\yearofpublication{2016}
\referencenumber{0.01}
\dateofsubmission{None}
\dateofrevision{None}
\issue{00}
\label{begin: 0}
\label{end: 0}
\else
\documentclass[12pt]{iopart}
\fi
\usepackage[colorinlistoftodos]{todonotes}
\usepackage{upgreek}
\usepackage{dcolumn}% Align table columns on decimal point

\newcommand{\zOn}{\ensuremath z_{\mathrm{On}}}
\newcommand{\zOff}{\ensuremath z_{\mathrm{Off}}}
\newcommand{\IOn}{\ensuremath I_{\mathrm{On}}}
\newcommand{\IOff}{\ensuremath I_{\mathrm{Off}}}
\newcommand{\BLf}{\ensuremath (Bl)_{\mathrm{F}}} 
\newcommand{\BLv}{\ensuremath (Bl)_{\mathrm{V}}}

\newcommand{\Vup}{\ensuremath U_{\mathrm{up}}}
\newcommand{\vup}{\ensuremath v_{\mathrm{up}}}
\newcommand{\Vdn}{\ensuremath U_{\mathrm{dn}}}
\newcommand{\vdn}{\ensuremath v_{\mathrm{dn}}}

\newcommand{\BarI}{\ensuremath \bar{I}}
\newcommand{\DeltaI}{\ensuremath \Delta I}
\newcommand{\mt}{\ensuremath m_{\mathrm{t}}}

\newcommand{\ds}{\ensuremath{\displaystyle}}

\newcommand{\ustat}{3.0}			%Statistical Uncertainty
\newcommand{\uel}{6.2}				%Electrical Uncertainty
\newcommand{\umag}{1.8}
\newcommand{\hres}{6.626\,069\,934(89)}
\newcommand{\hrel}{163}
\newcommand{\hrunc}{13}
\newcommand{\hchange}{15}  % 163.4 - 148

\begin{document}
\ifdouble
\title{Measurement of the Planck constant at the National Institute of
Standards and Technology from 2015 to 2017}
\else
\title[$h$ measured at NIST]{Measurement of the Planck constant at the National Institute of
Standards and Technology from 2015 to 2017}
\fi
\author{D.~Haddad$^1$, F.~Seifert$^{1,2}$, L.S.~Chao$^1$, A. Possolo$^1$, D.B.~Newell$^1$, J.R.~Pratt$^1$, C.J.~Williams$^{1,2},$ 
S.~Schlamminger$^1$}
%\email{darine.haddad@nist.gov}
%\affiliation{
%$^1$National Institute of Standards and Technology (NIST), 100 Bureau Drive Stop 8171, Gaithersburg, MD 20899, USA \\
%$^2$University of Maryland, Joint Quantum Institute, College Park, MD 20742, USA \\}
%}
\address{
$^1$National Institute of Standards and Technology (NIST),
         100 Bureau Drive Stop 8171,  Gaithersburg, MD 20899, USA \\
$^2$University of Maryland, Joint Quantum Institute, College Park, MD 20742, USA \\}    
%\ead{darine.haddad@nist.gov} 

\begin{abstract}

Researchers at the National Institute of Standards and Technology(NIST) have measured the value of the Planck constant to be  $h =\hres\times 10^{-34}\,$J\,s (relative   standard uncertainty $\hrunc\times 10^{-9}$). The result is based
  on over 10\,000 weighings of masses with nominal values ranging   from 0.5\,kg to 2\,kg  with the Kibble   balance NIST-4. The uncertainty  has been reduced by more than twofold relative to a previous  determination because of three factors:  
  (1) a much larger data set than previously available, allowing a more realistic, and smaller, Type A evaluation; 
  (2) a more comprehensive measurement of the back action of the weighing current on the magnet by weighing masses up to 2\,kg, decreasing the uncertainty associated with magnet non-linearity;
  (3) a rigorous investigation of the dependence of the geometric factor on the coil velocity reducing the uncertainty assigned to time-dependent leakage of current in the coil.
\end{abstract}

\maketitle{}
\section{Introduction}

This article summarizes measurements that were carried out with the Kibble balance, NIST-4, at the National Institute of Standards and Technology (NIST) from December  22, 2015 to April 30, 2017. A detailed description of NIST-4 and a first determination of the Planck constant $h$ with a relative standard uncertainty of $34\times 10^{-9}$ can be found in~\cite{Haddad2016}. Since the previous result,  several improvements to NIST-4 have been made. More importantly, many careful measurements and systematic investigations have improved our understanding of the apparatus, leading  to smaller estimates of three dominating uncertainties.

\section{The theory of the Kibble balance}

The principle of the Kibble balance, formerly known as watt balance, was first published by Bryan Kibble,~\cite{Kibble1976} a metrologist at the National Physical Laboratory in the United Kingdom. This section introduces the theory necessary to understand the improvements that led to the new result presented here, but it does not contain the complete theory of the Kibble balance. A comprehensive discussion of the principle of the Kibble balance can be found in~\cite{Robinson2016,Steiner2013}. The Kibble balance has a long history at NIST~\cite{Olsen1980,Olsen1989,Williams1998, Steiner2005a,Steiner2005b,Steiner2007,Schlamminger2014,Schlamminger2015} and the designation NIST-4 indicates that this is the fourth instrument that has been built and operated by researchers at NIST. Throughout the world several Kibble balances are being constructed or operated~\cite{Sanchez2014b,Thomas2015,Fang2014,Sutton2014,Kim2014}.

Common to NIST-1 through NIST-4 is that a wheel is used for both the balancing and moving mechanisms. The wheel pivots about a knife edge collinear with the wheel's central axis. A measurement coil and test mass are suspended from one side of the wheel while a tare mass is suspended from the other via multi-filament bands. The tare mass includes a small motor consisting of a coil in a permanent magnet system, similar in design but much smaller than the main magnet, for generating a force to rotate the wheel. The benefit of a wheel versus a traditional balance beam is that the former prescribes a pure vertical motion for the suspended coil whereas the latter traces an arc.

The measurement is performed in two modes: force and velocity mode. In force mode, a current $I$ in a coil with a wire length $l$ immersed in a radial magnetic field with magnetic flux density $B$ is controlled such that the balance wheel remains at a constant angle chosen  by the operator. While the balance wheel is servo controlled, a mass standard with  a mass $m$,  typically 1\,kg, can be placed on or removed from the mass pan. Without the mass standard on the pan, the electromagnetic force balances the excess mass on the tare side $\mt$ (usually about  $m/2$):
\begin{equation}
\IOff \BLf = \mt g. \label{eq:Foff}
\end{equation}
Here, $g$ denotes the local acceleration of gravity and $\BLf$ is the geometric factor of the magnet and coil combination, the product of $B$ and $l$ as measured in force mode. The current in the coil for the mass-off state is denoted by $\IOff$. With the mass standard on the mass pan, the current reverses to $\IOn$ and the force equation is
\begin{equation}
\IOn \BLf - m g = \mt g.\label{eq:Fon}
\end{equation}
Subtracting equation~\ref{eq:Fon} from~\ref{eq:Foff} and solving for $\BLf$ yields
\begin{equation}
\BLf  = \frac{\ds m g}{\IOn-\IOff} .
\end{equation}

Generally, the best results can be achieved by symmetrizing the measurement and the instrument as much as possible. Specifically, this is achieved by adjusting $\mt = m/2$, which results in the two equal but opposite currents. The advantage of using symmetric currents is explained in detail in section~\ref{sec:mag}.

The geometric factor is also obtained in velocity mode, where the coil is swept through the magnetic field by rotating the wheel with  the tare-side motor. During the coil sweep, the induced voltage $U$ and the coil's velocity $v$ are measured simultaneously. These coil sweeps can move either downward, with negative velocity $(\vdn<0$) or upward, with positive velocity ($\vup>0$). The geometric factor in the velocity mode is determined by
\begin{equation}
\BLv  = \frac{1}{2}\left( \frac{\Vup}{\vup} + \frac{\Vdn}{\vdn}  \right) .
\end{equation}
The up and down measurements are necessary to cancel small thermal and other parasitic voltages present in the circuit. These voltages are approximately a few hundred nanovolts. By averaging the measured geometric factors for  up and down sweeps, these extra voltages cancel, as long as they remain constant over the duration of the sweeps and $\vup=-\vdn$.

The ratio of the two measurements of the geometric factor $\BLf$ and $\BLv$ is nominally one. The mechanical quantities are measured in the International System of Units (SI), whereas the electrical quantities are measured in conventional units, denoted by the subscript $_\mathrm{90}$, hence 
\begin{equation}
\frac{\BLf}{\BLv}=\frac{ \left\{ \BLf \right\}_{\frac{\mathrm{N}}{\mathrm{A_{90}}}}}{\left\{\BLv\right\}_{\frac{\mathrm{V_{90}\,s} }{\mathrm{m} }} }\frac{ \mathrm{N\,m\,s}^{-1}}{\mathrm{ V_{90}\,A_{90}}} =
\frac{ \left\{ \BLf \right\}_{\frac{\mathrm{N}}{\mathrm{A_{90}}}}}{\left\{\BLv\right\}_{\frac{\mathrm{V_{90}\,s} }{\mathrm{m} }} }\frac{ \mathrm{W}}{\mathrm{ W_{90}}}.\label{eq:watt}
\end{equation}
The terms in the numerator and denominator of the ratio are  written as products of numerical quantities and units. The numerical quantity is indicated by the curly brackets $\left\{\right\}$ in the units given by the subscript. The last term of Equation~\ref{eq:watt} is a ratio of watts expressed in the International System of Units (SI) and conventional units. The ratio must be equal to one, since both measurements are determining the same physical quantity, the geometric factor. Hence,
\begin{equation}
\frac{ \left\{ \BLf \right\}_{\frac{\mathrm{N}}{\mathrm{A_{90}}}}}{\left\{\BLv\right\}_{\frac{\mathrm{V_{90}\,s} }{\mathrm{m} }} } =
\frac{ \mathrm{W_{90}}}{\mathrm{ W}}.
\label{eq:blratio}
\end{equation}

The value one can be written as the Planck constant divided by the Planck constant. Expanding the numerator as the product of a numerical quantity and the SI-unit and the denominator as a numerical quantity and the conventional unit yields
\begin{equation}
1=\frac{\left\{h\right\}_{\mathrm{W\,s}^{2}}}{\left\{h\right\}_{\mathrm{W_{90}\,s}^2}} 
\frac{\mathrm{W\,s}^{2}}{\mathrm{W_{90}\,s}^2} \;\;\mbox{and thus} \;
\frac{\left\{h\right\}_{\mathrm{W\,s}^{2}}}{\left\{h\right\}_{\mathrm{W_{90}\,s}^2}} = 
\frac{\mathrm{W}_{90}}{\mathrm{W}}.
\label{eq:hratio}
\end{equation}
By combining equations~\ref{eq:blratio} and \ref{eq:hratio}, an equation for the numerical value of the Planck constant can be obtained,
\begin{equation}
\frac{ \left\{ h     \right\}_\mathrm{SI}}
     {\left\{ h \right\}_{90}} = 
\frac{ \left\{ \BLf \right\} } 
     { \left\{ \BLv \right\} } .\label{eq:h}
\end{equation}
In equation~\ref{eq:h}, the expressions $\left\{ h \right\}_\mathrm{SI}$  and $\left\{ h \right\}_{90}$ are the numerical values of the Planck constant in SI and in conventional units, respectively. To define the conventional units, the numerical values of the conventional Josephson constant and the conventional von Klitzing constant were fixed in 1990~\cite{Taylor1989}. From these numerical values, the numerical value of the Planck constant in the conventional unit system can be obtained,
\begin{equation}
\left\{h\right\}_{90} = 6.626\,068\,854\,361\ldots\times 10^{-34}.
\end{equation}

\section{Overview of the data}
\label{sec:data}

\begin{figure}[htb]
\begin{center}
\includegraphics[width=\columnwidth]{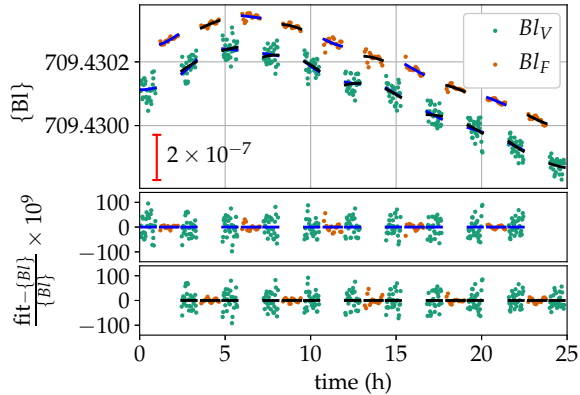}
\end{center}
\caption{A data run started on November 15 2016. The data is typical for NIST-4. Groups of measurements in velocity and force mode are carried out in an alternating pattern. The scatter in a velocity group is several times the scatter of the data in force mode. The blue and black line segments are second order polynomials that are fitted to three groups of data (velocity, force, velocity). The inner velocity groups are part of two fits (one shown in black, the other in blue). Their respective relative residuals are shown in the two plots below the main panel.
\label{fig:onerun}
}
\end{figure}

The measurements are organized in runs typically lasting about a day
each. A run usually comprises ten sets of determinations of the
geometric factors. Figure~\ref{fig:onerun} shows the measured geometric factors in force and velocity modes for a typical run. A set consists of three groups of measurements, two velocity groups and one force group. In each velocity group the coil is swept 30 times through the magnetic field in alternating directions (down, up, down, etc.). Each force group contains 17 weighings, alternating nine with the mass on the balance pan and eight with the mass off the balance pan.

 Figure~\ref{fig:onerun} shows the data collected in a typical run that was started at 15:21  local time on November 15, 2016. The run was terminated at 16:10 the next day by the operator and yielded ten data sets.   
One measurement  of the Planck constant is derived from the data in each set. The measured value is obtained from the difference in the zeroth order term of a second degree polynomial fitted separately to the data obtained in velocity mode and in force mode. The blue and black segments in Figure~\ref{fig:onerun} show the polynomial fits to every other set. Each velocity group, other than the very first and very last group is used for two adjacent $h$ measurements. The residuals of the polynomial fits are shown in the lower two graphs of Figure~\ref{fig:onerun}.

\begin{figure}[htb]
\begin{center}
\includegraphics[width=\columnwidth]{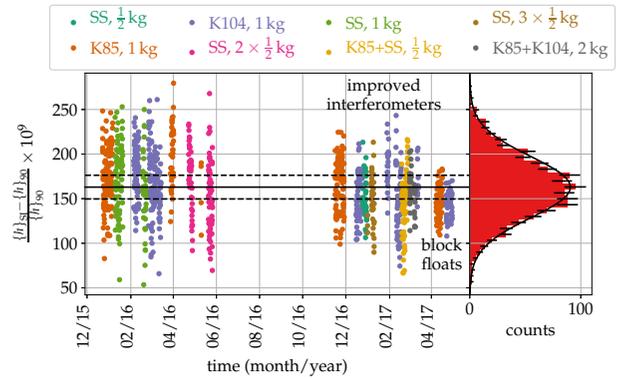}
\end{center}
\caption{The complete data set used for this determination of the Planck constant.\label{fig:alldata} Data taken with stainless steel masses are abbreviated as SS, the Platinum-Iridium prototypes are designated as K85 and K104. The solid horizontal line  indicates the final measurement result and the two dashed lines are drawn $\pm \hrunc \times 10^{-9}$  away from the result.}
\end{figure}

\begin{figure}[htb]
\begin{center}
\includegraphics[width=\columnwidth]{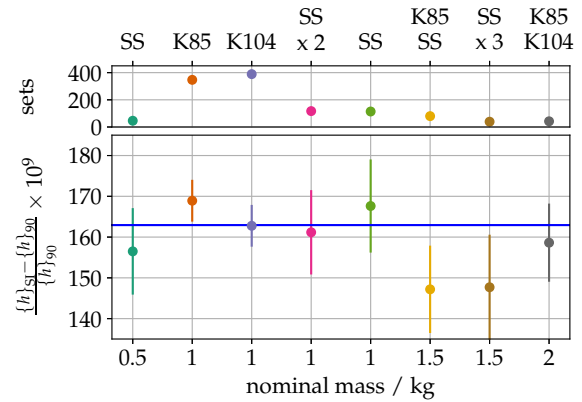}
\end{center}
\caption{The lower graph shows values of $h$ as a function of mass. A total of eight masses or mass combinations were used. The lower horizontal axis shows the nominal value, while the upper horizontal axis indicates the stacking that was used to obtain these values. The abbreviation SS stands for stainless steel and K85 and K104 denote two Pt-Ir prototypes. The upper graph shows the number of sets that were obtained for each mass or mass combination.
\label{fig:h-vs-m}}
\end{figure}

For the data discussed in this paper, a total of 1174 sets were measured from December 2015 through April 2017. Figure~\ref{fig:alldata} shows the $h$ measurements for all sets. A total of eight different combinations of masses were used. To combine the mass values we used a total of  five stainless steel (SS) masses with a nominal value of 0.5\,kg, one stainless steel mass with a nominal value of 1\,kg, and two Platinum-Iridium prototypes, labelled K85 and K104. Figure~\ref{fig:h-vs-m} shows the average value of $h$ for each mass combination used in the experiment. The top diagram in the figure shows the number of sets that were measured for each mass combination. The majority of the data were obtained using K85 and K104 with 347 and 389 sets, respectively.
  
The relative standard deviation of the velocity  and force residuals shown in Figure~\ref{fig:onerun} are $35\times 10^{-9}$ and $12\times 10^{-9}$. The residuals in velocity mode were improved during the summer 2016. The relative standard deviation of the residuals in one velocity group ranged from $30\times 10^{-9}$ to $60\times 10^{-9}$ before  and from $23\times 10^{-9}$ to $50\times 10^{-9}$ after summer 2016.  A second improvement was achieved at the end of March 2017 by installing a vibration isolation system. Eight air springs and a commercially available pneumatic controller position lift the concrete block supporting NIST-4 by 10\,mm off the building's foundation while maintaining its pitch and roll angle  to within few $\upmu$rad. Floating the block on air springs reduced the vibrational excitation of NIST-4 significantly. With the block floated, the relative standard deviation of the residuals in velocity mode are in the range from $11\times 10^{-9}$ to $25\times 10^{-9}$. Interestingly, the residuals of the fits to each individual volt-velocity profile improved by an order of magnitude, while the standard deviation of the residuals in velocity mode only decreased by a factor two. We assume that, after floating the block, the standard deviation of the residuals is limited by the variability from one sweep to the next.

The first reduction in the scatter of the residuals, over the summer of 2016,  was achieved by substantially stiffening the base plates and optimizing the mounting technique of the three interferometers used to  measure the velocity of the coil.  Before that time, each interferometer was screwed to the base plate of the Kibble balance. Three mounting plates made from $25.4\,$mm thick aluminium, each supporting one interferometer and turning mirrors, were mated to the base plate through kinematic mounts. This thicker plate and improved mounting to the Kibble balance decreased vibrational coupling, parasitic motions, and internal contortions  of the interferometers which led to the visible reduction of the scatter in Figure~\ref{fig:alldata}.  No data were included from the end of May 2016 to the beginning of November 2016, even though some data were collected during this period. The work was focused on improving the statistical uncertainty and not on collecting science data.

The measurements of $\BLf$ contain a measured value of the local acceleration of gravity $g$ using an absolute gravimeter. For the majority of the data presented here, the absolute gravimeter was operated simultaneously with the Kibble balance. Commercially available software was used to calculate the time dependent part of $g$ and added to the last measured value. The output of the software has been verified by using long data sets (several months) obtained with the absolute gravimeter in the Kibble balance laboratory. The effects included in the calculation of $g$ are tidal effects of the sun and moon, ocean loading, effects due to atmospheric pressure, and the effect of polar motion. The value of $g$ at the test mass centre is tied from the absolute reference in the laboratory and corrected for the vertical gradient of $g$~\cite{Leaman2015,Seifert2016}. The vertical gravity gradient was measured three different times at the mass pan location. Other than being corrected for $g$, the data shown in Figure~\ref{fig:onerun} is obtained from raw voltmeter readings and the voltage setting of the Programmable Josephson Voltage Standard. For the interferometer readings, the Abbe offset is considered when combining the three interferometers. 

\begin{figure}[htb]
\begin{center}
\includegraphics[width=\columnwidth]{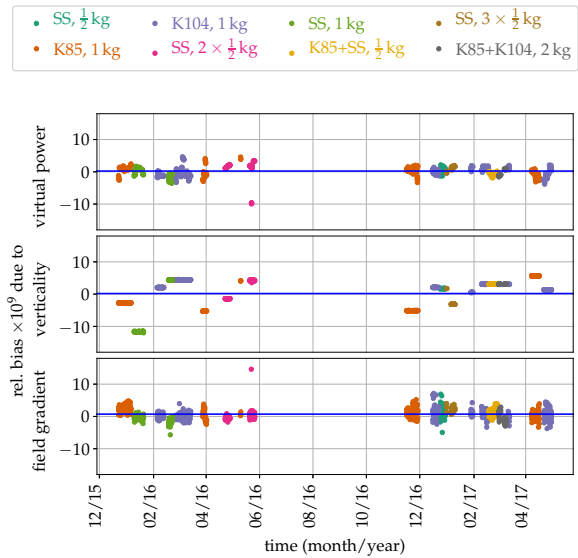}
\end{center}
\caption{
The alignment biases that have to be subtracted from the raw data of each set in order to obtain the values shown in Fig.~\ref{fig:alldata}. The horizontal blue lines show the averages of the biases. The three average values of the relative biases are between $0$ and $1 \times 10^{-9}$.  Data taken with stainless steel masses are abbreviated as SS, the Platinum-Iridium prototypes are designated as K85 and K104.}
\label{fig:biases}
\end{figure}

The raw measurements of the geometric factors in velocity and force mode contain a number of biases that need to be subtracted to obtain the final values of $h$. Every time a  bias is subtracted from the data, an uncertainty is added to the result because the bias is not precisely known.

For NIST-4, three different categories of biases need to be considered: (1) the back action of the current in the coil during force mode on the magnetic field; (2) diffraction and wavefront distortion in the measurement of the coil velocity with the three interferometers; (3) the alignment of the balance, the coil and the interferometers. 
The biases caused by the magnetic fields are discussed in detail below. 
Comprehensive information on the alignment biases can be found in~\cite{Haddad2016} and~\cite{Gillespie1997}. In brief, these alignment biases can be divided into three groups
\begin{description}
  \item[Virtual power] contains the sum of five relative parasitic power terms. These are the products of non-vertical forces and torques on the coil in force mode with non-vertical velocities and angular velocities in velocity mode. For example the term  $F_x\,v_x/(F_z\,v_z)$ is the parasitic power in the $x$ - direction. The other four terms are translation in $y$ and rotations around the $x$, $y$, and $z$ axes. A discussion of these terms is given in~\cite{Gillespie1997}.
  \item[Verticality] collects four terms that arise from slight misalignment of the three interferometer beams with respect to $g$. Three of the four terms are the result of a parasitic motion of the coil in velocity mode. A translation  along $x$, $y$, and a rotation about $z$ of the coil can attenuate or augment the measured vertical velocity of the coil: a perfect vertical laser beam is insensitive to a displacement perpendicular to the direction of the laser beam. But if the laser beam is slightly misaligned, then a parasitic horizontal displacement will have a component along the direction of the laser beam. The fourth term is independent of the coil's parasitic motion and reflects the fact that if a measurement beam deviates from verticality by $\alpha$, the measured velocity is attenuated and will be $\cos{(\alpha)}$ times the vertical velocity. 
  \item[Field gradient] consists of four terms that capture the relative difference in the geometric factor between velocity mode and force mode. The coil position in force mode is not exactly  on the trajectory of the coil in velocity mode and a correction must be applied. Therefore, the differences of the coil position in force mode to the coil trajectory in  $x$, $y$, $\theta_x$, and $\theta_y$ are multiplied by  the measured derivatives of the geometric factors with respect to these directions.
  \end{description}

Figure~\ref{fig:biases} shows the relative correction that needs to be subtracted from the raw data as a result of the three types of alignment biases. The blue horizontal lines in Figure~\ref{fig:biases} are the values obtained by averaging the corrections over all data sets. The averages are  $0.2\times 10^{-9}$, $0.2\times 10^{-9}$, and $0.7\times 10^{-9}$ for the biases caused by virtual power, verticality, and field gradients, respectively. The averages are very small and overall the biases have a negligible effect (relatively at most $1.1\times 10^{-9}$) on the reported result. 

Figure~\ref{fig:biases} also shows how often the alignment was checked and the apparatus realigned. For example, it can be seen that the verticality of the interferometers was measured almost every time the mass in the experiment was changed.

\section{Improved understanding of the apparatus}
The relative standard uncertainty of the result given here is less than half of that published previously~\cite{Haddad2016}. This improvement is due to smaller uncertainties in three categories of the uncertainty budget (Table~\ref{tab:uncert-budget}) labelled statistical, magnetic field, and electrical. In the following three sections, the new uncertainty estimates for these three categories are discussed.

\subsection{Estimation of the statistical uncertainty}

In the 2016 publication, the relative statistical uncertainty was estimated to be $24.9\times 10^{-9}$, obtained from the standard deviation of the 125 determinations that were incorporated into the corresponding estimate of $h$. This assigned uncertainty was very conservative because the uncertainty associated with an average is generally smaller than the uncertainty associated with the   observations that are averaged. If $n$ observations, each with the same uncertainty, are uncorrelated, then their average will have an uncertainty that is $\sqrt{n}$ smaller than their common, individual uncertainty. However, since the 2016 estimate was based on only 13 days of data, it was not quite possible to detect and characterize any pattern of correlations reliably. Furthermore, given the level of familiarity with the instrument at the time, it was difficult to ascertain whether the series of measured values was stationary. Both the inability to gauge auto-correlations meaningfully and the lack of clarity regarding stationarity led us to adopt a rather conservative assessment of this component of uncertainty.

  The current measurement of $h$ is based on data acquired over the course of about 16 months which produced 1174 individual determinations of $h$, depicted in Figure~\ref{fig:alldata}. During this long period, the Kibble balance underwent several mechanical upgrades. The multi-filament band  connecting the main coil to the wheel was replaced by a similar band with higher tensile strength. As stated earlier, the three interferometers were remounted on thicker, kinematically mounted base plates. Additional optics were installed or replaced in the interferometers to better measure the laser beam verticality, reduce frequency leakage, and measure the parasitic coil motion. Eight different masses or combinations of masses were employed for this measurement duration. Finally, the instrument was completely realigned on several occasions. Still, the resulting values of $h$ remained essentially constant during this period (Figure~\ref{fig:alldata}).
  
To determine whether the standard error of an average of $n$ consecutive observations is inversely proportional to $\sqrt{n}$, we undertook a sub-sampling analysis similar to~\cite{Mosteller}. Refer to~\cite{Politis1999} for a detailed description of sub-sampling methods in general. The procedure is as follows: 
\begin{description}
%\begin{enumerate}[(a)]
\item[(a)] The 1174 data points, sorted by their time stamp, are
    partitioned into $m_{k}$ blocks, each of which comprises $k$
    consecutive observations. Only $k m_{k}$ observations are used and
    the left over $1174 - k m_{k}$ are discarded (for this particular
    value of $k$).
\item[(b)] The $m_{k}$ block averages are computed and the standard
    deviation $s_{k}$ of these averages is calculated.
%\end{enumerate}
\end{description}
If the observations were uncorrelated and all had the same mean and
standard deviation $\sigma$, then $s_{k}$ should be close to
$\sigma/\sqrt{k}$ (of course, as $k$ increases, the number $m_{k}$ of
blocks of size $k$ decreases and the relationship is
increasingly obfuscated by sampling noise). 

For positively correlated observations, the variability of the block averages will be greater than $\sigma/\sqrt{k}$  and one may claim it is as if the blocks had not $k$ but some $k_{\mathrm{EFF}} < k$ observations
each where $k_{\mathrm{EFF}}$ is an \emph{effective sample size} that
incorporates a ``correction'' for the presence of positive serial
correlation. This effective sample size can be estimated based on
values of the ratio $s_{k}^{2}/(s^{2}/k)$, for different values of
$k$, where $s$ denotes the standard deviation of the complete
dataset. Examination of these ratios for increasing values of $k$
reveals that, for our data, $k/k_{\mathrm{EFF}}$ approaches 8.75. Hence,
the effective sample size is $134 \approx 1174/8.75$, and the relative
contribution from this source of measurement uncertainty is estimated
to be $\ustat\times 10^{-9} = s/\sqrt{134}$, where $s = 34.3\times 10^{-9}$ is the standard deviation of the complete data set.

The analysis described above was carried out using conventional averages and standard deviations. Using robust analogs of both, as described in~\cite{Huber81}, resulted in essentially the same estimate of the effective sample size.

\subsection{Dependence of the geometric factor on the weighing current}
\label{sec:mag}

In force mode, the coil is carrying a current while in the velocity mode current is absent. This difference challenges a fundamental assumption of the Kibble balance experiment: the geometric factor is the same in both modes. The current in the coil generates a magnetic field thereby altering the state of the permanent magnet system. Thus, there is reason to believe that the geometric factor in force mode differs slightly from the one in velocity mode. The geometric factor in the presence of weighing current is usually parametrized as
\begin{equation}
Bl(I) = Bl_o (1 + \alpha I + \beta I^2), \label{eq:BLI}
\end{equation}
where $Bl_o$ is the unperturbed geometric factor and $\alpha$ and $\beta$ are two specific parameters that depend on the detailed design of the magnet system~\cite{Robinson2016}. These denote the relative sensitivity of the magnet system to the weighing current and its squared value. 

A second effect that can cause a current dependent deviation from the ideal Kibble balance theory is the reluctance effect~\cite{Schlamminger2013}. The energy of a current carrying coil is given by $E=-\frac{1}{2} L I^2$, where $L$ denotes the self-inductance of the coil. If the self-inductance of the coil depends on the vertical position of the coil, a force $F = \frac{1}{2} I^2 L^\prime$ acts on the coil. Here, $L^\prime = \mbox{d}L/\mbox{d}z$ denotes the first derivative of the self-inductance of the coil with respect to the vertical coil position. The force is pointing towards the location where the coil's self-inductance is maximal, typically the centre of the magnet. For example, a solenoid actuator is based on this effect where a soft iron slug is pulled into a solenoidal coil after the coil is energized with either direct or alternating current. 

To properly take into account the dependence of the geometric factor on the current and the reluctance force, the simplified equations~\ref{eq:Foff} and~\ref{eq:Fon} have to be amended to read
\begin{equation}
\IOff Bl(\IOff) + \frac{1}{2} \IOff^2 L^\prime(\zOff) =  \mt g   \label{eq:nbloff}
\end{equation}
and
\begin{equation}
\IOn Bl(\IOn) + \frac{1}{2} \IOn^2  L^\prime(\zOn)  -m g =   \mt g  \label{eq:nblon}.
\end{equation}
As indicated in  equations~(\ref{eq:nbloff}) and (\ref{eq:nblon}), the derivative of the self-inductance with respect to the vertical position has to be taken at the coil positions, $\zOff$ and $\zOn$, corresponding to the weighing positions, mass off and mass on, respectively. In force mode, the balance is servo controlled to a fixed position where the  mass exchange occurs. But due to the finite stiffness of the coil suspension, the vertical coil positions  for mass on and off differ slightly, about $13\,\upmu$m per 1\,kg. The difference in the $z$ position must also be accounted for in the calculation of the $Bl$. A correction for the different weighing positions can easily be applied since the velocity mode measurements reveal the profile of $Bl$ as a function of $z$. This detail is left out in the discussion below.

The equations simplify considerably if the mean values ($\bar{\;\;}$) and the differences ($\Delta$) to the means are used for positions and currents, i.e.,
\begin{equation}
\bar{z}=\frac{\zOn+\zOff}{2} \mbox{,}\;\;  \Delta z=\frac{ \zOn-\zOff}{2}\label{eq:z}
\end{equation}
and
\begin{equation}
\BarI  =\frac{\IOn+\IOff}{2} \mbox{,}\;\;  \DeltaI = \frac{\IOn-\IOff}{2}. \label{eq:I}
\end{equation}
In normal operation, the currents are generated with magnitudes equal but opposite to each other, hence $\Delta I$ equals about $\IOn$, which is 6.9\,mA for a 1\,kg  mass standard. On the other hand, $\BarI$ is small, usually less than 7\,$\upmu$A. The size of $\BarI$ can be adjusted by adding or removing small tare masses from the suspended parts on either side of the wheel. 

To calculate the first derivative of the self-inductance at the two coil positions, a Taylor series expansion is used
\begin{equation}
L^\prime(\bar{z} \pm \Delta z) \approx L^\prime (\bar{z}) \pm \Delta z L^{\prime\prime} (\bar{z}).\label{eq:L}
\end{equation}
As mentioned above, the difference in coil position for the two weighing states is due to elasticity in the coil suspension, hence $\Delta z$ can be parametrized as
\begin{equation}
\Delta z = \frac{\Delta F}{\kappa} \approx \frac{Bl_o \DeltaI}{\kappa},\label{eq:dz}
\end{equation}
where $\kappa$ denotes the spring constant of the mechanical system. For NIST-4, $\kappa = 0.7\,\mbox{N}/\upmu\mbox{m}$. 
Combining equations~\ref{eq:BLI} %\ref{eq:nbloff}, \ref{eq:nblon}, \ref{eq:z}, \ref{eq:I}, \ref{eq:L}, and 
through~\ref{eq:dz} and solving for the force yields
\begin{equation}
mg = 2 \Delta I Bl_o (1-c_\mathrm{mag}),
\end{equation}
where $c_\mathrm{mag}$ is a correction term due to the effects caused by the weighing current. This term is given by
\begin{eqnarray}
c_\mathrm{mag} &\approx& -c_1 \BarI -c_2 \BarI^2 -c_3\DeltaI^2 \;\;\mbox{with} \label{eq:magc} \\
c_1&=&  2 \alpha + L^\prime(\bar{z})/Bl_o, \\
c_2&=&  3 \beta + \frac{1}{2}\,L^{\prime\prime}(\bar{z} )/\kappa, \mbox{ and}\\
c_3&=& \beta +  \frac{1}{2}\,L^{\prime\prime}(\bar{z})/\kappa.
\end{eqnarray}

%\begin{eqnarray}
%c_\mathrm{mag} \approx &-&\BarI     \left( 2 \alpha + L^\prime(\bar{z})/Bl_o \right) \nonumber \\ 
%                 &-&\BarI^2   \left( 3 \beta + \frac{1}{2}\,L^{\prime\prime}(\bar{z} )/\kappa\right) \nonumber   \\ 
%                 &-&\DeltaI^2 \left( \beta +  \frac{1}{2}\,L^{\prime\prime}(\bar{z})/\kappa \right).\label{eq:magc}
%\end{eqnarray}
The correction is organized in three components that are proportional to $\BarI$, $\BarI^2$, and $\DeltaI^2$ -- no term proportional to $\DeltaI$ arises from this theory. The terms proportional to $\BarI$ and $\BarI^2$ can be made arbitrarily small by reducing the mass imbalance by adding or removing masses on the counter mass side until the absolute values of the currents match perfectly. The adjustment is done in air, so buoyancy effects need to be taken into account in order to achieve $\BarI=0$ in vacuum. Besides a few runs that were used to determine the sensitivity of the result on $\BarI$,  the absolute value of $\BarI$ was below $6\,\upmu$A.

In contrast to $\BarI$, $\DeltaI$ can not be reduced and is given by the mass that is used in the experiment. Hence the term that is proportional to $\DeltaI^2$, needs to be precisely determined and applied as a correction to the measured data.

The focus throughout the 2016/17 measurement campaign was to obtain a better value for the magnetic effect.  The shortcoming of the previous determination of this effect was that only a 1.5\,kg mass was used, resulting in an effect that is only 2.25 times that of 1\,kg mass standard. Hence, in  the 2016/17 measurement campaign, data were gathered using mass values ranging from 0.5\,kg to 2\,kg in 0.5\,kg steps. The 2\,kg value was achieved by stacking two Platinum-Iridium prototypes, K85 and K104, on top of each other. In this situation, the quadratic effect of the weighing current is quadrupled compared to a measurement with a 1\,kg mass standard. But, no change in the measured $h$ value was seen. The data quality obtained with the two prototypes was very high and thus a good limit could be placed on this effect.

These measurements yield
\begin{eqnarray}
c_1 &=& (4.608 \pm 0.003)\times 10^{-6}\,\mbox{mA}^{-1}, \nonumber\\
c_2 &=& (1.0 \pm 0.5)\times 10^{-11}\,\mbox{mA}^{-2},\mbox{ and}\nonumber\\
c_3 &=& (0.03 \pm 0.03)\times 10^{-11}\,\mbox{mA}^{-2}.\nonumber
\end{eqnarray}
From these values $\beta$ and $L^{\prime\prime}$ can be obtained:
\begin{eqnarray}
\beta 			&=& (0.50 \pm 0.23)\times 10^{-9}\,\mbox{mA}^{-2},\mbox{ and}\nonumber\\
L^{\prime\prime} &=& (-656 \pm 332)\,\mbox{H}/\mbox{m}^2.\nonumber
\end{eqnarray}
Before the magnet was installed in NIST-4, the second derivative of the self inductance of a  coil in the magnet with respect to its position was measured to be $L^{\prime\prime} = -346\,$H/m$^2$~\cite{Seifert2014}, which agrees with the measurement obtained here within one standard uncertainty. For the measurement in~\cite{Seifert2014}, the coil used was different, but had a similar number of turns. 
%The number obtained here is consistent with an estimate in~\cite{Schlamminger2013}.

For a 1 kg mass standard ($\DeltaI=6.9\,$mA) the correction given by $c_3\DeltaI^2$ is $(1.4\pm1.4)\times 10^{-9}$. The correction agrees with zero within one standard deviation. Averaged over the 1174 measured sets the uncertainty of the correction is $1.7\times 10^{-9}$. The slightly higher uncertainty is caused by the measurements of masses and mass combinations with mass values greater than 1\,kg.

For the result published in~\cite{Haddad2016}, a determination of $\beta$ was made by measuring three different mass values: 0.5\,kg, 1\,kg, and 1.5\,kg. The resulting relative bias on the final result due to $\DeltaI^2$ was estimated to be $(17.5\pm 15.4)\times 10^{-9}$ for a 1 kg mass standard. The determination presented here has a smaller uncertainty and supersedes the previous determination of this bias. The improvement was possible by employing higher mass values (2\,kg) and by obtaining data with better quality.

\subsection{The effect of time-dependent leakage}

The effect of electrical leakage is a concern in Kibble balance experiments.  A discussion of these effects can be found in~\cite{Robinson2012e}. Interestingly, a pure resistive leakage path across the coil does not affect the result. For NIST-4, this cancellation was verified by placing a $R_p=100\,\mbox{M}\Omega$ resistor parallel to the coil with $R_c=112\,\Omega$. The result obtained with the $100\,\mbox{M}\Omega$ parallel to the coil did not differ from the result without this resistor within the relative measurement uncertainty of $30\times 10^{-9}.$

The measurement is only independent of the size of the leakage resistance if the system is completely linear, i.e, described by ideal circuit elements. Two types of non-linearities can limit this cancellation and can give rise to a systematic effect: if the leakage resistance is  voltage or time-dependent. In normal operation of NIST-4, a 1\,kg mass standard is used and the coil is moved with a nominal velocity of $v_{\mathrm{nom}}=975\,\upmu \mbox{m}\,\mbox{s}^{-1}$. For these parameters, the voltages across the coil are almost the same in both modes with 0.68\,V and 0.69\,V in velocity and force mode, respectively.

\begin{figure}[htb]
\begin{center}
\includegraphics[width=\columnwidth]{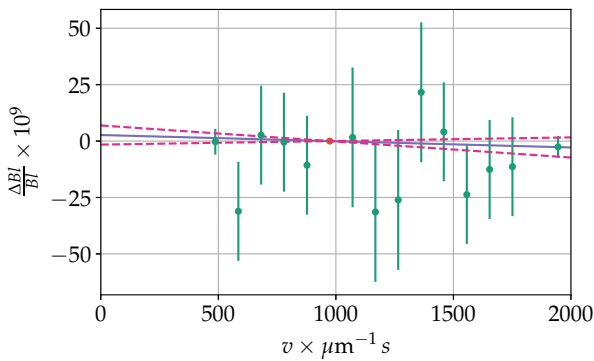}
\end{center}
\caption{Measurements of the relative difference of the geometric factor measured at velocity $v$ from the geometric factor measured at the nominal velocity. The error bars denote the 1-$\sigma$ statistical uncertainties associated with the measurements. The solid line is a least squares fit with a constraint to pass through the point of nominal velocity and $\Delta B=0$. The one sigma  uncertainty interval of the fit is given by the dashed lines surrounding the best fit.}
\label{fig:bl-vs-velo}
\end{figure}

A time-dependent leakage resistance remains a concern for the NIST-4 measurements. The possible effect of the time-dependent storage of charge can be assessed by measuring the geometric factor in velocity mode with different velocities of the coil. Changing the velocity changes the timing, in particular how long the voltage is applied to various parts of the circuit. For example, 
the science data discussed in Section~\ref{sec:data} was taken with a nominal coil velocity of $v_{\mathrm{nom}}=975\,\upmu \mbox{m}\,\mbox{s}^{-1}$. The total length of the sweep, including acceleration and deceleration is 78\,mm. Hence, the voltage is applied on the coil for 40\,s before the coil reaches the weighing position. Reducing this velocity by a factor of two extends the travel time by a factor of two.

To estimate the effect of the time-dependent leakage on the measurement, we took a group of 30 sweeps with a velocity $v\neq v_{\mathrm{nom}}$ bracketed by two groups of 30 sweeps with $v_{\mathrm{nom}}$ on either side. The analysis of this set (= three groups) is done similarly to the data analysis of the science data i.e., two parallel second degree polynomials were fitted to the measurements. The difference of the polynomials in $Bl$ is the result of one measurement. Many measurements were taken at 13 different velocities ranging from $v_{\mathrm{nom}}/2$ to $2v_{\mathrm{nom}}$. Most of the measurements were concentrated on the smallest and largest velocity. Figure~\ref{fig:bl-vs-velo} shows the obtained differences in $\Delta Bl =Bl(v)-Bl(v_{\mathrm{vnom}})$. A least squares adjustment of the data to a straight line, $\Delta Bl/Bl = \gamma (v-v_{\mathrm{vnom}})$ yields $\gamma= (-2.76 \pm 4.35)\times 10^{-9}\,\mbox{mm}^{-1}\,\mbox{s}$. 

This result agrees with zero within its uncertainty and, thus, no correction to the data was applied. The relative uncertainty in $h$ due to this effect for measurements carried out with a nominal velocity of $975\,\upmu \mbox{m}\,\mbox{s}^{-1}$ is $4.24\times 10^{-9}$. Previously a value of $10\times 10^{-9}$ was used for this line in the uncertainty budget. This larger value was estimated from experiences with NIST-3. The time dependent leakage is a part of the electrical uncertainty. The relative uncertainty of this category reduces from $10.9\times 10^{-9}$ to $\uel\times 10^{-9}$.

\section{Result and Discussion}
Analysing 1174 sets taken between December 2015 and April 2017, a final result of the Planck constant,
\begin{equation}
h =  \hres\times 10^{-34}\,\mbox{J}\,\mbox{s},
\end{equation}
is obtained. This value corresponds to 
\begin{equation}
\frac{ \left\{h\right\}_\mathrm{SI}-\left\{h\right\}_{90}} { \left\{h\right\}_{90} } = ( \hrel \pm \hrunc )\times 10^{-9}
\end{equation}

The relative standard uncertainty of this result is $\hrunc\times 10^{-9}$. The main categories of the uncertainty budget are listed in Table~\ref{tab:uncert-budget}. For comparison, the uncertainties that were assigned for 2016 publication~\cite{Haddad2016} are also listed.

\newcommand{\sss}{\hspace{1em}}
\begin{table*}[h!]
\begin{center}
%\begin{tabular}{ll}
\begin{tabular}{lrrrr}
%\begin{tabular}{lD{.}{.}{1}D{.}{.}{1}}

 & \multicolumn{2}{c}{this measurement} & \multicolumn{2}{c}{previous measurement}\\
Source & \multicolumn{1}{c}{item}  &\multicolumn{1}{c}{ category} &\multicolumn{1}{c}{item}&\multicolumn{1}{c}{category}\\ 
       & \multicolumn{1}{c}{$ u/h\times 10^9$ } &\multicolumn{1}{c}{$ u/h\times 10^9$} 
       & \multicolumn{1}{c}{$ u/h\times 10^9$ } &\multicolumn{1}{c}{$ u/h\times
       10^9$} \\
\hline\hline
\sss  Calibration of resistor			    & 4.5  & & 4.5 \\
\sss  Time dependent leakage			    & 4.2  & & 10.0 \\
\sss  Leakage in velocity mode			    & 0.7  & & 0.7  \\
\sss  Leakage in force mode					& 0.5  & & 0.5 \\
\sss  Josephson Voltage standard		    & 0.3  & & 0.3  \\
\sss  Grounding								& 0.0  & & 0.0  \\
\bf Electrical 								&	   & \bf \uel & & \bf 10.9\\
\hline
\sss  Calibration of mass			    	& 5.7 & & 5.5  \\
\sss  Transport						    	& 2.0 & & 0.0  \\
\sss  Sorption						    	& 0.3 & & 0.3  \\
\sss  Magnetic effects				    	& 0.3 & & 3.0 \\
\bf Mass metrology							&   & \bf 6.1 & & \bf 6.3\\
\hline
\bf Profile fitting							&	& \bf 5.0 & & \bf 5.0 \\
\hline
\bf Balance mechanics						&	& \bf 5.0 & & \bf 5.0		\\
\hline
\sss  Laser verticality						& 4.3 & & 5.4\\
\sss  Field gradient						& 1.5 & & 2.3 \\
\sss  Virtual power							& 1.2 & & 2.7\\
\sss  Abbe Offset							& 0.1 & & 0.8 \\
\bf Alignment								&	& \bf 4.7	& & \bf 6.5\\
\hline
\sss  Statistical							& 2.5 & & 2.5\\
\sss  Site 									& 2.1 & & 2.1 \\
\sss  Water table							& 2.0 & & 2.0 \\
\sss  Instrument 							& 1.6 & & 1.6\\
\sss  Tie 									& 1.0 & & 1.0\\
\sss  Vertical translation					& 0.6& & 0.6\\
\sss  Additional corrections				& 0.2 \\
\bf Local acceleration, $g$					&	& \bf 4.3 & &  \bf 4.3\\
\hline
\bf Statistical								&	& \bf \ustat & & \bf 24.9\\
\hline
\sss  Corrections for $\Delta I^2$			&1.7&& 15.4\\
\sss  Corrections for $\BarI$				&0.4&& 0.4 \\
\sss  Corrections for $\BarI^2$				&0.2&& 0.2 \\

\bf Magnetic field							&	& \bf \umag & & \bf 15.4\\
\hline
\sss  Jitters in photo receivers 			& 1.2 & & 1.2\\
\sss  Synchronization      					& 1.0 & & 1.0\\
\sss  Diffraction							& 0.6 & & 0.6\\
\sss  Frequency leakage    					& 0.4 & & 0.4\\
\sss  Wavelength		    				& 0.0 & & 0.0\\
\sss  Beam shear							& 0.0 & & 0.0\\
\sss  Time interval analyser timing 		& 0.0 & & 0.0\\
\bf Velocity								&	& \bf 1.7 & & \bf 1.7		\\
\hline\hline
 \bf Total relative uncertainty	& & \bf 13.5 & &\bf 33.6\\
\end{tabular}
\end{center}
\caption{Sources of uncertainty and their relative magnitudes for measurements of $h$ with the Kibble balance NIST-4. All entries are relative standard uncertainties $(k=1)$. Entries with $0.0$ denote uncertainties that are smaller than $0.05\times 10^{-9}$. Column two and three indicate the uncertainties in the present measurement and column four and five indicate the uncertainties in the previous measurement~\cite{Haddad2016}.
The lines in bold are categories which may consist of several individual items printed in regular font above the category. The categories as well as the items within are sorted by size of the uncertainty in the present measurement. }
\label{tab:uncert-budget}
\end{table*}

The new result includes the data that were used to measure $h$ in 2016~\cite{Haddad2016}. Superseding the 2016 value, the value reported here is relatively larger by $\hchange\times 10^{-9}$. One reason for this increase is that the previous value included a relative correction of $17.5\times 10^{-9}$ due to a change in the geometric factor in response to the symmetric part of the weighing current. A more precise study showed that the change is only about $1.4\times 10^{-9}$ for a 1 kg mass standard.

The smaller uncertainty of the new measurement is due to the reduction of three line items in the uncertainty budget.
\begin{description}
\item[1] A new assessment of the statistical uncertainty that was made possible by the larger data set gave an estimate of $\ustat\times 10^{-9}$, about eight times smaller than the original estimate of $24.9\times 10^{-9}$. 
\item[2] A careful measurement of the influence of the weighing current on the geometric factor reduced the uncertainty due to the magnetic field from $15.4\times 10^{-9}$ to $1.8\times 10^{-9}$. 
\item[3] A detailed measurement of the geometric factor in velocity mode with different coil velocities constrained the size of a possible time-dependent leakage effect. This reduced the contribution of the uncertainty category electrical from $10.9\times 10^{-9}$ to $\uel\times 10^{-9}$.
\end{description}

The uncertainty budget discussed above is for the measurement of the Planck constant where it is possible to accumulate a large data set. After the revision of the International System of Units, NIST-4 will be used to realize the mass unit. In this case, a mass value has to be measured much more quickly. A conservative estimate gives a statistical uncertainty of $21.8\times 10^{-9}$ for a 24 hour long measurement. This yields a total relative standard uncertainty of $25\times 10^{-9}$. Integrating four days of data will reduce the total relative uncertainty to below $20\times 10^{-9}$.

\section*{Acknowledgements}
Several precise measurements are necessary to determine the Planck constant with a Kibble balance and it seems impossible for a small research group to know all the details necessary to succeed. Fortunately, the greatest asset of the National Institute of Standards and Technology are its researchers and technicians who are not only experts in their respective subject matter, but are also passionate to share, teach, and apply their expertise to improve measurements in all areas of metrology. 

We would like to thank Zeina Kubarych, Patrick Abbott, and Edward Mulhern for many mass calibrations, very often on short notice. We appreciate the work of Rand Elmquist, Marlin Kraft, and Dean Jarrett to provide traceability to the quantum Hall resistance standard and advice on all matters relating to DC resistance metrology. We are grateful to Sam Benz, Charlie Burroughs, Alain R\"ufnacht, Yi-hua Tang, Stefan Cular, and Jason Underwood for their help with the programmable Josephson voltage standard.

%\section*{References}

\end{document}